\documentclass {article}
\usepackage{epsfig}

\begin{document}

\title{Observation of individual molecules trapped on~a~nanostructured insulator}

\date{}

\maketitle

\begin{center}
\large{L. Nony$^{1}$\footnote{to whom correspondence should be addressed
(E-mail~: L.nony@unibas.ch)}, E. Gnecco$^{1}$, A. Baratoff$^{1}$, A. Alkauskas$^{1}$,
\and R.~Bennewitz$^{2}$, O. Pfeiffer$^{3}$, S. Maier$^{3}$, A. Wetzel$^{3}$, E. Meyer$^{1, 3}$, and~Ch. Gerber$^{1, 4}$}
\end{center}

\noindent
\small{$^1$ National Center of Competence in Research "Nanoscale Science", University of~Basel,
Klingelbergstrasse 82, CH-4056 Basel, Switzerland}\\
\small{$^2$ Dept.~of Physics, McGill University, 3600 University street, Montreal PQ, H3A 2T8 Canada}\\
\small{$^3$ Institute of Physics, Klingelbergstrasse 82,
CH-4056 Basel, Switzerland}\\
\small{$^4$ IBM Research Division, Zurich Research Laboratory, CH-8803 R\"uschlikon, Switzerland}

\textbf{Published in Nanoletters 4(11), 2185-2189 (2004)}

\begin{abstract}
For the first time, ordered polar molecules confined in
monolayer-deep rectangular pits produced on an alkali halide
surface by electron irradiation have been resolved at room
temperature by non-contact atomic force microscopy. Molecules self-assemble in a specific fashion inside pits of width smaller than 15~nm.
By contrast no ordered aggregates of molecules are observed on flat terraces.
Conclusions regarding nucleation and ordering mechanisms are drawn. Trapping in pits as small as 2~nm opens a route to address
single molecules.
\end{abstract}

\vspace{12pt}

The adsorption of medium-sized ($<100$ atoms) functional organic
molecules on patterned surfaces has become the subject of
intensive studies motivated by the prospect of hybrid molecular
electronic devices \cite{joachim00a}. The operation of such
devices should be governed by the electronic properties of a single
or, at most, of a few of the constituent molecules. With this goal
in mind, numerous groups have used scanning tunneling
microscopy (STM) to study a variety of organic molecules adsorbed primarily on metal surfaces
\cite{umbach98a,langlais99a, dewild02a, barlow03a, rosei03a, berner03a}.
Besides atomic-scale spatial resolution, STM provides the means
for an \textquotedblleft accurate placement of molecules in appropriate position
and orientation to form a device\textquotedblright \cite{joachim00a}. However, as
discussed in ref.\cite{joachim00a}, many hurdles remain on the way
towards the development and the integration of useful molecular
electronic devices. In particular, it is desirable to electrically isolate
the device; this can be achieved using an insulating substrate. The extension of atomic-scale characterization and manipulation studies to surfaces of insulators is therefore of high interest. It is, however, quite
challenging because STM can no longer be applied, unless one deals
with ultrathin insulating films on metals as substrates \cite{schintke01a}.
A more fundamental motivation for such studies is to understand
the delicate balance between intermolecular and molecule-substrate
interactions which determines the mobility and aggregation of
molecules, as well as their eventual ordering on ionic substrates.

Since 1995 non-contact atomic force microscopy (nc-AFM),
has proven capable of yielding images with atomic resolution on a
wide variety of insulating surfaces \cite{morita02a}.
This technique relies on a micro-fabricated tip at the end of a
cantilever excited at its fundamental bending eigenfrequency
\cite{giessibl03a}. Upon approaching the surface, the tip first
senses an attractive force which decreases the resonance
frequency. The negative frequency shift, $\Delta f$, varies
rapidly with the minimum tip-distance, especially a few $\mbox{\AA }$
away from the surface. When $\Delta f$ is used for distance control,
contrasts down to the atomic scale can be achieved in ultrahigh
vacuum.

Recently, the technique has been used to study molecules adsorbed
on various surfaces \cite{note01}. But so far, only few results
have been reported concerning molecules adsorbed intact on
insulators, presumably owing to the high mobility of such molecules at room
temperature \cite{nony04a}. Thus, in order to obtain ordered
molecular arrangements with a definite orientation on insulating
surfaces, it is mandatory to find means to lower the mobility of
the molecules. In the present work, we report a nc-AFM study of
chloro [subphthalocyaninato] boron(III) (SubPc) molecules confined
on a modified KBr(001) surface in ultra-high vacuum (UHV) at room
temperature. For the first time, an ordered arrangement of
molecules has been directly observed on a nanostructured insulator
consisting of intentionally created rectangular pits acting as
molecular traps.\\

\begin{figure}
 \includegraphics[width=7cm]{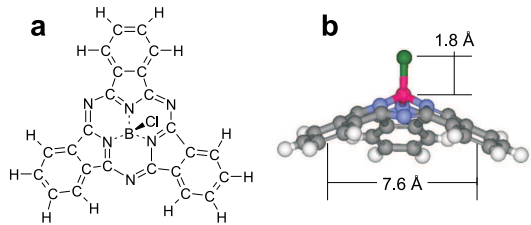}
 \caption{SubPc molecule: (a) Schematic top view of the chemical structure; (b) Side
 view, with characteristic dimensions  and atomic species (Cl: green, B: pink, N: blue, C: gray, H: white;
 from ref.~\cite{berner03a}).}
 \label{FIGSUBPC}
\end{figure}

Cleaved alkali halide single crystals have been among the first insulating
surfaces showing atomic periodicity and resolution of atomic-scale
defects in nc-AFM experiments \cite{note04}. Clean flat surfaces
can be rather easily prepared and imaged in UHV. Straight edged pits can be created by exposing the surface to an
electron beam for a short time \cite{kolodziej01}. The pits are
square or rectangular and one monolayer (ML),
\emph{i.e.} 0.33~nm, deep on KBr(001). First evidence for confinement, albeit without
molecular resolution was found for flat PTCDA molecules \cite{nony04a}.
Here, polar cone-shaped molecules are used. The chemical
composition and structure of SubPc are shown in
figs.~\ref{FIGSUBPC}(a, b). The central boron atom is sp$^3$
coordinated to an apex chlorine atom and to three isoindole
moieties. The properties and potential applications of SubPc and
related molecules have recently been reviewed \cite{claessens02}.
The choice of this molecule has initially been motivated by its large dipole moment (5.4~D)
pointing away from the protruding electronegative Cl atom which might form a strong ionic bond to substrate cations.
Density functional theory (DFT) calculations predict a dipole moment close to the measured one
and an
electrostatically fitted effective charge of $\sim - 0.4e$ on~Cl~\cite{berner03a}.

The dynamic force microscope is a home-built instrument placed in
a UHV chamber with a base pressure of about $5 \times
10^{-11}$~mbar \cite{bennewitz00a}. We used a silicon cantilever
with a supersharp tip \cite{Nanosensors} (resonance frequency
$f_0=165,513$~Hz, nominal spring constant $k=48$~N/m, quality
factor $Q=30,800$). During the experiments, the tip is oscillating
with a constant amplitude $A\simeq14$~nm ($\pm10\%$ calibration uncertainty).

The KBr single crystal was cleaved in air, transferred to the vacuum
chamber, heated to $393$~K and irradiated with the $1$~keV
electron beam of a LEED electron source to produce the pits
\cite{bennewitz02a}. SubPc molecules were sublimed in UHV from a
resistively heated aluminum oxide crucible while the sample was
kept at $353$~K and the pressure below $10^{-10}$~mbar. A
quartz-microbalance was used to calibrate the deposition rate. One
ML of molecules was deposited, with an estimated error of about
$20\%$ \cite{note06}.

\begin{figure}
 \includegraphics[width=8cm]{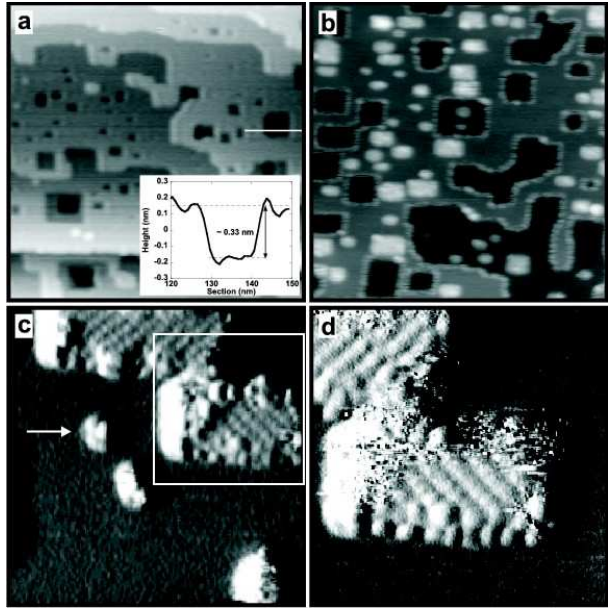}
 \caption{(a) Non-contact AFM image of the KBr(001) surface after electron irradiation (frame edge: 150~nm,
 $\Delta f = -10$~Hz). Monatomic steps and pits, one layer deep, are visible. The bar indicates
 the location of the section shown in the inset. (b, c, d)- Non-contact AFM images of SubPc molecules on the
 electron-irradiated KBr(001) surface. (b) Frame size:~$100$~nm, $\Delta f=~-5$~Hz. The molecules decorate steps and fill pits of width smaller than $15$~nm. (c) Frame size:~$27$~nm, $\Delta f = -8$~Hz. No molecules are observed on the KBr terraces.
 Molecular self-assembled structures are revealed within two pits. The arrow indicates a 2~nm-wide pit trapping a small
 number of molecules ($4$ or $5$). The square denotes the scanning area shown in (d). (d) Frame size:~$18$~nm, $\Delta f = -8$~Hz. Molecular resolution reveals
 that the structures are tilted $\pm 45^o$ with respect to the pits edges. Images (c, d) were processed
 using a technique which enhances the molecular contrast.} \label{FIGKBRSUBPC}
\end{figure}

Figure~\ref{FIGKBRSUBPC}(a) shows a nc-AFM image of the KBr(001)
surface at a constant frequency shift, $\Delta f=-10$~Hz, after
irradiation, but before deposition of the molecules. Monatomic
steps and regular pits are clearly visible. Atomic resolution was
obtained on flat (001) terraces at more negative $\Delta f$; as
expected, only one type of ionic species was imaged as maxima
\cite{bennewitz02a}. Figures~\ref{FIGKBRSUBPC}(b, c, d) show
images of the sample after deposition of 1~ML SubPc. Molecules and
substrate ions could not both be resolved with the same $\Delta
f$, likely due to different force \emph{vs.} distance
characteristics \cite{note04}. The wide-area image in
fig.~\ref{FIGKBRSUBPC}(b) provides an overview of the way the
molecules are distributed. They decorate steps and edges of wide
pits, but only pits with a size smaller than $15$~nm are filled
(rectangular bright areas). No molecules are present on the flat
terraces. In fig.~\ref{FIGKBRSUBPC}(c), maxima attributed to
single molecules can be recognized inside two pits and along their
edges. The arrow indicates a $2$~nm-wide pit trapping only $4$ or
$5$ molecules. This is consistent with the nucleation mechanism
described hereafter. In fig.~\ref{FIGKBRSUBPC}(d), the trapped
molecules appear self-assembled into regular rows oriented $+45^o$
(upper pit) or $-45^o$ (central pit) with respect to the edges.
The distance between adjacent rows is $1.4$~nm and the distance
between two consecutive molecules in a row is $1.0$~nm. The
apparent height of the rows is about $0.6$~nm with respect to the
surrounding KBr(001) terrace, or equivalently $0.9$~nm with
respect to the bottom of the pit. The molecules are rather
well-ordered in the center of each pits, but the structure appears
somewhat mismatched along their edges. A blurred bright area
attributed to mobile molecules and a few protruding molecules are
also visible on top of the organized layer. When attempting to
image at a more negative $\Delta f$, thereby decreasing the
tip-sample distance, horizontal stripes appeared, implying that
material was dragged by the tip while scanning.\\

\begin{figure}
 \includegraphics[width=8cm]{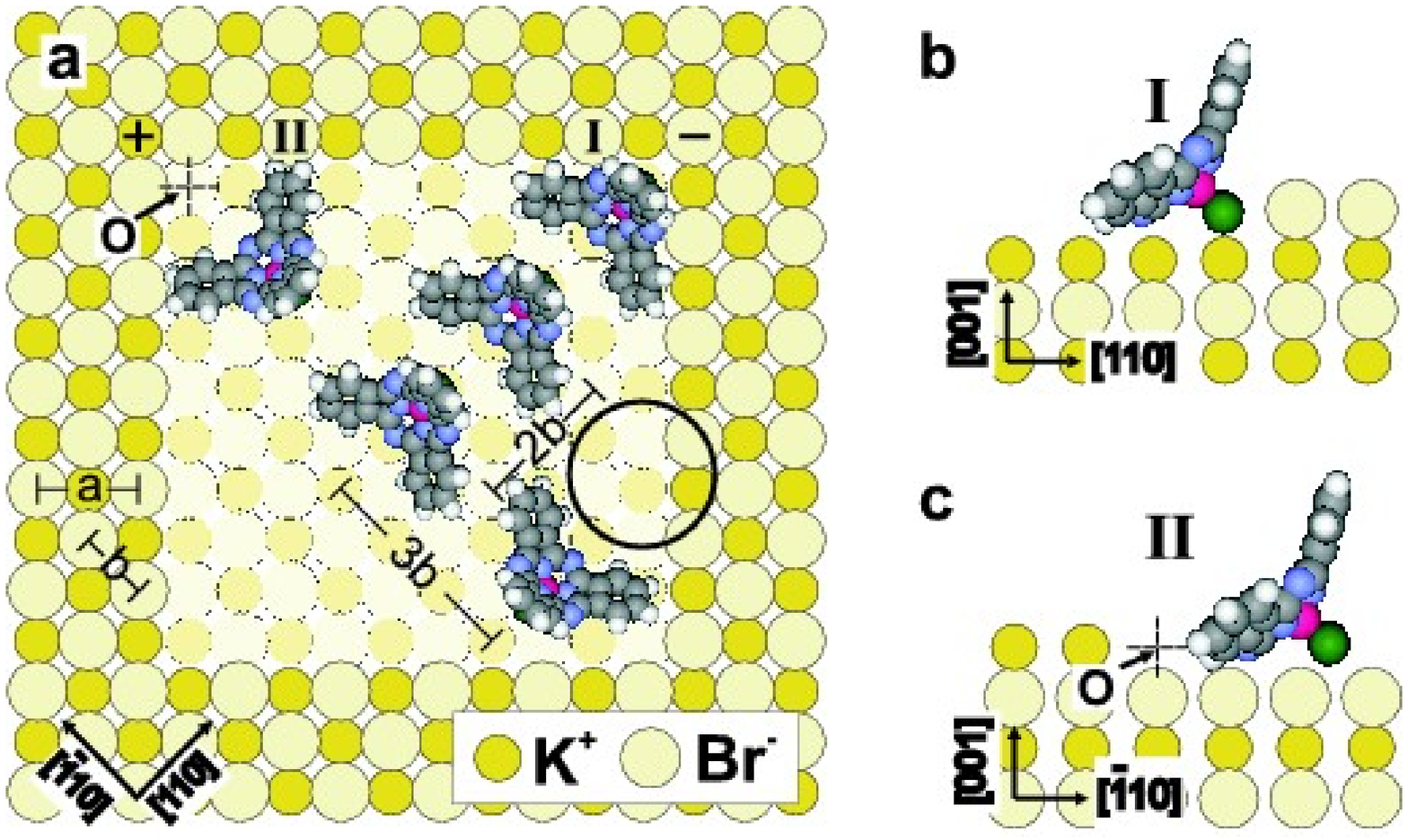}
 \caption{(a) Possible schematic arrangements of SubPc molecules inside a
   pit consisting of $10\times10\times1$ missing ions. The lattice constant of KBr is $a=0.66$~nm, thus $b=a/\sqrt 2=0.467$~nm. The distance between two molecules in the diagonal row is
   $2\times b\simeq 1$~nm, while the distance between rows is $3\times b\simeq 1.4$~nm, in accordance with the experimental
    data. The Cl apex atom is strongly attracted to the electrostatic potential well near the corner site (position I),
    resulting  in a tilt of the molecular axis. Position II is
    energetically unfavorable.  (b) and (c) Side view along the [110] and [$\overline{1}$10] directions showing the
    approximate orientation of the molecule in position I and II, respectively.}
  \label{FIGSUBARRANGEMENT}
\end{figure}

Let us consider the possible arrangement of the molecules inside
the pits. Figure~\ref{FIGSUBARRANGEMENT}(a) shows that the
alignment of SubPc's in rows tilted $\pm 45^o$ with respect to the
[100] direction of the KBr(001) surface can match the underlying
lattice if the distance between two adjacent rows is $3\times
b=1.40$~nm and if two consecutive molecules in a row are separated
by $2 \times b = 0.93$~nm, in good agreement with the
experimentally observed spacings. However, this argument alone
cannot explain why molecules are only found trapped within the
smallest pits and not on the terraces. To address this question,
we first discuss interactions of a single SubPc molecule with the substrate
and then those between molecules. Intermolecular bonds are strong and no new covalent bonds are
expected to form. We therefore treat each SubPc molecule as rigid
and interacting only via electrostatic and Van der Waals plus
steric interactions. The latter are described by Lennard-Jones
6-12 pair potentials, using UFF parameters and combination rules
\cite{Rappe92}.

\begin{figure}
 \includegraphics[width=10cm]{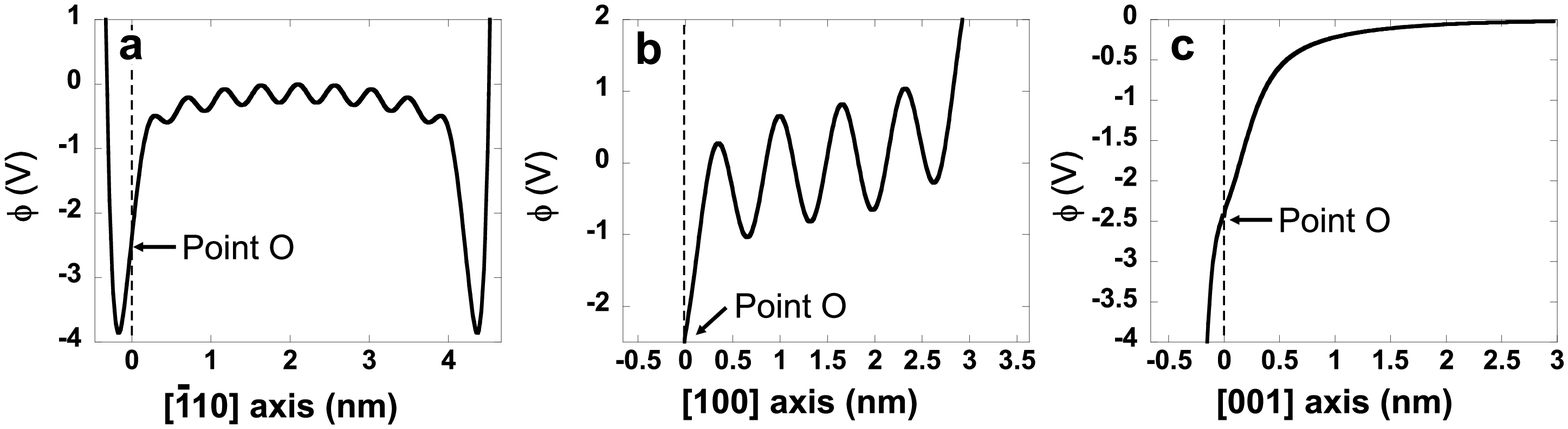}
 \caption{Electrostatic potential inside the pit displayed in fig.~\ref{FIGSUBARRANGEMENT}(a)
  along lines passing through point O in (a) the [$\overline{1}$10], (b) the [100] and (c) the [001] directions, respectively.
  The potential was calculated by summing up the Coulomb contributions of $100\times100\times100$
  alternating electrical charges $\pm e$ at the unrelaxed positions of the ions and subtracting the contribution
  of a $10\times10\times1$ squared monolayer slab in the center.}
 \label{FIGPOTENTIAL}
\end{figure}

Steps and pit edges expose ions of alternating sign, thereby
preserving overall charge neutrality. The resulting electrostatic
potential $\phi$ therefore oscillates parallel to those edges with
a period equal to the lattice constant $a = 0.66$~nm, as it does
along [100] and [010] directions above a flat terrace.  Laplace's
equation implies that this oscillating component decays
exponentially in empty space. The ions exposed at the pit corners
(see fig.~\ref{FIGSUBARRANGEMENT}(a)) induce an additional
component which varies on the scale of the pit width, except close
to the corners. These features are readily apparent in
figs.~\ref{FIGPOTENTIAL}(a-c) which display $\phi$ computed along
[$\overline{1}$10], [100] and [001] directions passing through the
first missing  K$^+$ ion in the upper left corner (point O in
fig.~\ref{FIGSUBARRANGEMENT}(a)).  The potential was computed for
an array of point charges $\pm e$ at unrelaxed positions of the
ions, an approximation which is sufficient for the following
semi-quantitative discussion.  Note in particular the pronounced
minimum (actually a 3D saddle point) in fig.~\ref{FIGPOTENTIAL}(a)
caused by the three Br$^-$ ions next to point O.

The rapid variation of $\phi$ over molecular dimensions requires
taking into account the charge distribution within a SubPc
molecule. In order to describe electrostatic interactions, it is
sufficient to approximate that distribution by partial atomic
charges fitted to the electrostatic potential of an isolated
molecule in a surrounding shell. So-called CHelpG charges \cite{breneman90}
were obtained from a density functional computation reported in
ref.\cite{berner03a}. The appreciable charge on the Cl atom
($-0.42~e$) leads to a strong attraction to the upper right corner
for a molecule in position I in
figs.~\ref{FIGSUBARRANGEMENT}(a,b), but is, however, mitigated
by the steric repulsion from the neighboring K$^+$ ions. Taking into
account appropriate UFF radii, we estimate an electrostatic energy
gain of about $1$~eV for the Cl atom. Owing to the decay of $\phi$
and to the alternating signs of the charges on the more distant B,
C and N atoms, their contribution is an order of magnitude
smaller.  Van der Waals attraction favors the adsorption of a
SubPc molecule to the bottom of the pit via two of its isoindole
\textquotedblleft feet\textquotedblright and also its trapping in position I. The resulting trapping
energy is so large compared to $k_BT$ at room temperature that it
prevents a SubPc molecule from diffusing away. Therefore corner
sites which expose K$^+$ ions can act as efficient nucleation
centers for molecules like SubPc. In the center of the pit,
however, the modulation of $\phi$ is only $0.3$~V.  The
corresponding corrugation of the potential energy surface is
dominated by the electrostatic interaction of the protruding Cl
atom and leads to a barrier of about $0.15$~eV, \emph{i.e.} almost
$6 k_BT$.  This is still insufficient to prevent diffusion of a
single molecule at room temperature over the time needed to image
one (about 12 scanlines, each lasting about $1$~s). Moreover, the
bias introduced by the slowly varying $\phi$ component, apparent
in figs.~\ref{FIGPOTENTIAL}(a,b), induces a drift of a molecule
adsorbed inside the pit towards position I or to the equivalent
site on the opposite side of the diagonal. This implies that
molecules trapped near opposite corners should be rotated by
$180^o$, but this prediction is difficult to verify.
Position II in
figs.~\ref{FIGSUBARRANGEMENT}(a,c) is much less favorable for
trapping because charges, or atoms, which contribute most to
electrostatic or Van der Waals attraction are kept away by steric
repulsion of the feet from the pit edges.

The previous estimates are qualitatively consistent with our
observations, namely that no molecules are imaged on flat
terraces, while they appear trapped along the edges of steps
and of pits smaller than $15$~nm. Besides, pits as small as $2$~nm
appear filled; judging from figs.~\ref{FIGPOTENTIAL}, this width
is sufficient for the required potential profile to develop. The
close packing along different diagonals inside the two pits
visible in fig.~\ref{FIGKBRSUBPC}(d) can be rationalized if those
pits expose ions of different sign at their bottom right corners.

In an attempt to understand the observed ordering, we calculated
the interaction between two SubPc molecules, using the same CHelpG
charges and UFF parameters.  The molecules were aligned as in the
row schematically depicted in fig.~\ref{FIGSUBARRANGEMENT}(a) with
their Cl atoms and two feet constrained to be in Van der Waals
contact with a plane. The intermolecular binding energy, which is
largely dominated by the Van der Waals component, is only $0.1$~eV
at the distance $2 b$, but increases by another $0.2$~eV if the
molecules are allowed to reach their equilibrium separation,
namely 7.25 \AA $\simeq 1.5 b$, \emph{i.e.} implying a serious
mismatch with respect to the substrate corrugation. Therefore the
observed stable commensurate order inside pits of intermediate
size is presumably the consequence of a complex interplay between
electrostatic and Van der Waals interactions between the
molecules, the bottom and the edges of the pit, which requires
more detailed calculations.

In any case, intermolecular interactions favors the relative
orientation of molecules in adjacent rows shown in
fig.~\ref{FIGSUBARRANGEMENT}(a). Note that the image in
fig.~\ref{FIGKBRSUBPC}(d) reveals irregularities, in particular
along the pit edges. Once a molecule is trapped at position I and
additional molecules arrange according to the supposed structure,
steric repulsion prevents perfect matching along the pit edges
(\emph{cf.} black circle in fig.~\ref{FIGSUBARRANGEMENT}(a)). This
mismatch must introduce some deviations from perfect ordering even
in the absence of irregularities in the pit structure itself.
Compared to fig.~\ref{FIGPOTENTIAL}(a), the larger modulation of
$\phi$ seen in fig.~\ref{FIGPOTENTIAL}(b) implies a sizable
diffusion barrier $\simeq 0.7$~eV along the pit edge.
This phenomenon, which is a specific property of steps on ionic
crystal surfaces \cite{nony04a}, might explain why molecules
trapped along extended steps, e.g. at the edges of wide pits,
appear poorly ordered.
\\

In conclusion, the nc-AFM measurements and theoretical estimates
indicate that molecules with a protruding atom or small moiety
carrying an appreciable charge (owing to electronegativity
differences between constituent atoms) can be trapped inside few
nanometer-wide pits created by electron irradiation on a KBr(001)
surface. Keeping in mind the goals outlined in the introduction,
the next challenge is to study the trapping of various molecules,
"molecular wires" in particular \cite{langlais99a,Yoon03},
equipped with judiciously positioned polar groups on different
nanostructured ionic substrates. In that connection, note that
monolayer deep pits have been created on several alkali halide
(001) surfaces and that their average number, size and separation
can be controlled by adjusting the electron dose and the substrate
temperature \cite{kolodziej01}. A further challenge would
obviously be to achieve intramolecular resolution. For that
purpose, beyond the difficulty of the immobilization of the
molecules, the tip preparation as well as the use of small
amplitudes \cite{Giessibl04} are two key issues that are currently
under investigation.

The authors acknowledge the Swiss National Center of Competence in
Research on Nanoscale Science and the National Science Foundation
for financial support. They thank S.~Berner (University of
Uppsala),
J.J. Kolodziej (Jagiellonian University, Krakow),
M.~von~Arx and
L.~Ramoino (University of Basel) for discussions and advice,
D.~Schlettwein and W.~Michaelis from University of Oldenburg for
providing the SubPc molecules, R.~Allenspach and R.~Schlittler
from IBM Zurich Research Laboratory for advice on the design of
the molecular evaporator.


\begin{thebibliography}{99}

\bibitem{joachim00a}
C. Joachim, J. Gimzewski, and A. Aviram, Nature \textbf{408}, 541 (2000).
\bibitem{umbach98a}
E. Umbach, K. Gl\"ocker, and M. Sokolowski, Surf. Sci. \textbf{402-404}, 20 (1998).
\bibitem{langlais99a}
V. Langlais, R. Schlitter, H. Tang, A. Gourdon, C. Joachim, and J. Gimzewski, Phys. Rev. Lett. \textbf{83}, 2809 (1999).
\bibitem{dewild02a}
M. de Wild, S. Berner, H. Suzuki, H. Yanagi, D. Schlettwein, S. Ivan, A. Baratoff, H.-J. G\"untherodt, and T. Jung, ChemPhysChem \textbf{3}, 825 (2002).
\bibitem{barlow03a}
S. Barlow and R. Raval, Surf. Sci. Repts. \textbf{50}, 201 (2003).
\bibitem{rosei03a}
F. Rosei, M. Schunack, Y. Naitoh, P. Jiang, A. Gourdon, E. Laegsgaard, I. Stensgaard, C. Joachim, and F. Besenbacher, Progr. Surf. Sci. \textbf{71}, 95 (2003).
\bibitem{berner03a}
S. Berner, M. de Wild, L. Ramoino, S. Ivan, A. Baratoff, H.-J. G\"untherodt, H. Suzuki, D. Schlettwein, and T. Jung, Phys. Rev. B \textbf{68}, 115410 (2003).
\bibitem{schintke01a}
S. Schnitke, S. Messerli, M. Pivetta, F. Patthey, L. Libioulle, M. Stengel, A. DeVita, and W.-D- Schneider, Phys. Rev. Lett. \textbf{87}, 276801 (2001).
\bibitem{morita02a}
S. Morita, R. Wiesendanger, and E. Meyer, \emph{Noncontact Atomic Force Microscopy} (Springer, Berlin, Germany, 2002).
\bibitem{giessibl03a}
F. Giessibl, Rev. Mod. Phys. \textbf{75}, 949 (2003).
\bibitem{note01}
\emph{cf.} for instance Ch. 11, 12 and 13 in ref. \cite{morita02a}
\bibitem{nony04a}
L. Nony, R. Bennewitz, O. Pfeiffer, E. Gnecco, A. Baratoff, E. Meyer, T. Eguchi, A. Gourdon, and C. Joachim, Nanotechnology \textbf{15}, S91 (2004).
\bibitem{note04}
\emph{cf.} Ch. 5 in ref. \cite{morita02a}
\bibitem{kolodziej01}
J. Kolodziej, B. Such, P. Czuba, F. Krok, P. Piatkowski, P. Strutski, M. Szymonski, R. Bennewitz, S. Sch\"ar, and E. Meyer, Surf. Sci. \textbf{482-485} (2001).
\bibitem{claessens02}
C.G. Claessens, D. Gonzalez-Rodriguez, and T. Torres, Chem. Rev. \textbf{102}, 835 (2002).
\bibitem{bennewitz00a}
R. Bennewitz, A. Foster, L. Kantorovich, M. Bammerlin, C. Loppacher, S. Sch\"ar, M. Guggisberg, E. Meyer, H.-J. G\"untherodt, and A. Shluger, Phys. Rev. B \textbf{62}, 2074 (2000).
\bibitem{Nanosensors}
Nanosensors, \emph{http://www.nanosensors.com}
\bibitem{bennewitz02a}
R. Bennewitz, O. Pfeiffer, S. Sch\"ar, V. Barwich, and E. Meyer, Appl. Surf. Sci. \textbf{188}, 232 (2002).
\bibitem{note06}
We used a deposition rate of 0.25 nm/min during 2 min. This coverage approximately corresponds to a full monolayer of the hexagonal close-packed superstructure of SubPc on Ag(111) reported in Ref. \cite{berner03a}. A proportional-integral controller implemented with LabView\texttrademark varies the current to keep constant the power input to the evaporator, thus giving a better reproducibility than deposition at constant current.
\bibitem{Rappe92}
A.K. Rapp\'{e}, C.J. Casewit, K.S. Colwell, W.A. Goddard III, and W.M. Skiff, J. Am. Chem. Soc. \textbf{114}, 10024 (1992).
\bibitem{breneman90}
C. Breneman and K.B. Wiberg, J. Comput. Chem. \textbf{11}, 361 (1990).
\bibitem{Yoon03}
D.H. Yoon, S.B. Lee, K.-H. Yoo, J. Kim, J.K. Lim, N. Aratani, A. Tsuda, A. Osuka, and D. Kim, J. Am. Chem. Soc. \textbf{125}, 11062 (2003).
\bibitem{Giessibl04}
F.J. Giessibl, S. Hembacher, M. Herz, Ch. Schiller, J. Mannhart,
Nanotechnology \textbf{15}, S79 (2004).
\end{thebibliography}
\end{document}